# A methodology to Evaluate the Usability of Security APIs

Chamila Wijayarathna and Nalin A.G. Arachchilage
School of Engineering and Information Technology
University of New South Wales
Australia
c.diwelwattagamage@student.unsw.edu.au and nalin.asanka@adfa.edu.au

*Abstract*—Increasing number of cyber-attacks demotivate people to use Information and Communication Technology (ICT) for industrial as well as day to day work. A main reason for the increasing number of cyber-attacks is mistakes that programmers make while developing software applications that are caused by usability issues exist in security Application Programming Interfaces (APIs). These mistakes make software vulnerable to cyber- attacks. In this paper, we attempt to take a step closer to solve this problem by proposing a methodology to evaluate the usability and identify usability issues exist in security APIs. By conducting a review of previous research, we identified 5 usability evaluation methodologies that have been proposed to evaluate the usability of general APIs and characteristics of those methodologies that would affect when using these methodologies to evaluate security APIs. Based on the findings, we propose a methodology to evaluate the usability of security APIs.

## I. INTRODUCTION

Information and Communication Technology (ICT) is a major aspect that researchers and industry are looking to leverage in their journey to sustainable development [1]. Even though ICT can be leveraged in many positive ways to make lives of people better in a sustainable way, a major problem that discourages the use of ICT is the increasing number of cyber attacks that we hear everyday. Despite the continuous evolution of security technologies, it appears that hackers are still capable of identifying security vulnerabilities of software applications to make attacks against computer systems. One of the main ways of vulnerabilities getting introduced into software applications is mistakes that application programmers make while they are developing applications [2].

Since most of the programmers who are involved in the software development process are not experts of cyber security and related aspects [2], they use Application Programming Interfaces (APIs) that provide security related functionalities to embed security functionalities to applications they develop [2]. These APIs that provide security related functionalities are known as security APIs [3], [4]. When security APIs that programmers use are not usable, it is difficult for programmers to learn and use APIs and hence, leads them to make mistakes that would result in introducing security vulnerabilities to applications they develop [4], [7], [30].

In an experiment conducted by Fahl et al. [8], they examined 13500 android applications and identified that 8% of those applications contained security vulnerabilities. They identified that mistakes made by programmers due to lack of usability of security APIs as a major reason for those vulnerabilities. In a similar experiment, Fahl et al. [9] revealed security vulnerabilities in iOS applications that have been caused by mistakes that programmers made while using security APIs that provide functionalities related to Secure Socket Layer (SSL) and transport Layer Security (TLS).

It has been acknowledged that one of the main reasons for the lack of usability in security APIs is that currently there is no proper methodology to evaluate the usability of security APIs [10]. Due to the lack of a proper usability evaluation methodology, usability issues of security APIs are left undiscovered and programmers who are using the security API will encounter those usability issues while using the API. Even though there have been few methods that are introduced to evaluate the usability of general APIs, there have been no investigation on the applicability of those methodologies in the security API context. If there is a methodology to evaluate the usability of security APIs, software developers who develop security APIs can evaluate the usability of security APIs they develop and identify usability issues exist in them. In that way, they can improve the usability of security APIs they deliver to be used by programmers, who are not experts of security and who are susceptible to make mistakes while using those APIs. Therefore, in this paper, we propose a methodology to evaluate the usability of security APIs.

We conducted a literature review on existing methodologies for evaluating the usability of general APIs. We identified 5 methodologies that have been proposed and used to evaluate the usability of general APIs. By reviewing literature, we attempted to identify strengths and weaknesses of each methodology when using them to evaluate the usability of security APIs. Based on these results, we propose one of those methodologies as the most suitable candidate to evaluate the usability and identify usability issues of security APIs.

The paper is structured as follows. In the next section, we present 5 usability evaluation methodologies (UEMs) that we identified by conducting the literature review. Then we identify qualities that a UEM should have in order to be a good candidate for evaluating the usability of security APIs and which UEM fulfills each quality. Then we present the methodology that we propose as the best candidate to evaluate the usability of security APIs. Finally we discuss the limitations of the proposed methodology and conclude the paper.

## II. USABILITY EVALUATION TECHNIQUES FOR GENERAL APIs

By conducting a systematic literature review through previous research that discuss "API usability", we identified five methodologies that have been presented for evaluating the usability of general APIs.

### A. Conducting user studies

User study based API usability evaluations are conducted with users of an API (i.e. programmers) [11]–[13]. In this method, programmers are recruited and given an individual task to complete that makes use of the API under evaluation. This task can be writing, reading or debugging a code that uses the API. Usability issues are identified by observing the programmer while completing the task or from the feedback programmers give after completing the task. Some studies have employed a think-aloud study [14] while participants are completing the task to identify problems they encounter.

Clarke [11] claims that Microsoft uses user studies with a Cognitive Dimensions Framework (CDF)-based questionnaire for evaluating the usability of APIs. In their methodology, once a participant completes the task, they must answer a questionnaire based on their experience in completing the task. Usability issues are identified by analyzing the answers to the questionnaire. Piccioni et al. [15] also used a similar methodology with a different set of questions to evaluate the usability of an API. There have been more variants of user studies that use different types of tasks and different methods for collecting feedback from participants [16]–[19].

Since user studies make use of programmers in the evaluation process, those will reveal actual issues that programmers experience while using the API to develop applications. Due to this, including users in the evaluation process is considered as the gold standard of usability engineering [12], [13], [20]. Involving programmers in the security API usability evaluation process will be important because security issues caused by the usability of security APIs occur when programmers incorrectly use security APIs. By getting them involved in the evaluation process, evaluators can understand what leads programmers to use the API incorrectly [12]. Furthermore, Petrie and Power [21] point out that user studies are much more likely to identify usability issues related to security. However, recruiting participants for user studies can be costly [12], [20]. Furthermore, user studies can be difficult to apply for APIs with a large number of classes and methods.

### B. Heuristic evaluation

Heuristic evaluation is a low cost, less time consuming method that can be used to evaluate large APIs as well as small APIs [20], [21]. In heuristic evaluation, an expert who is a knowledgeable person in usability and the application domain inspects the API according to a set of heuristics and identifies issues in it [20]. Grill et al. [20] conducted an API usability study, which used heuristic evaluation to identify usability issues of an API.

However, past studies show that issues identified in heuristic evaluation are different from actual issues that users/programmers experience [12], [20], [21]. Furthermore, it has been identified that heuristic evaluation gives more emphasis to less severe issues [12], [21] and it gives less emphasis to security related usability issues [21]. Even though heuristic evaluation has been often used to evaluate the usability of end user applications, there is not much evidence available on API usability studies that use heuristic evaluation other than work conducted by Grill et al. [20] and Mosqueira-Rey et al. [22].

### C. API Walkthrough method

The API Walkthrough method [23] is also conducted using participants who have profiles similar to real users of the API as in user studies. During the evaluation process, the facilitator will ask participants to walk through a code that uses the API and to explain the code according to their understanding. Participants can follow a think-aloud protocol while walking through the code [23]. Usability issues are then identified by analyzing the participants' code explanation and their think-aloud output. O'Callaghan [23] claims that this method is used successfully at Mathworks[1], for evaluating the usability of APIs they develop. Farooq and Zirkler [24] also proposed a similar method, which is known as the API peer review method.

The API walk-through method also makes use of programmers in the evaluation process. Furthermore, it can be used very early in the development process, even before implementation and documentation is started. Therefore, issues identified in the evaluation can be easily fixed in the design. This also allows the API walk-through method to be easily integrated to the software development life cycle. Furthermore, this method is less time consuming and requires fewer resources [23], [24]. However, Farooq and Zirkler [24] say that this method identifies less usability flows and does not identify issues related to some usability aspects. Furthermore, not considering API documentation is one of the major limitations of this method for evaluating the usability of security APIs. Issues in API documentation and user guides lead programmers to make assumptions that might be wrong, which may introduce vulnerabilities into the applications they develop.

### D. API concept maps method

Gerken et al. [25] introduced the API concept maps method to evaluate the usability of APIs. In this method, participants have to create a map to show the relationship between the API and their code (which can be a given task or a real application) within a 30-60 minute session. Each participant has to repeat this once a week for about 5 weeks and update the concept map that they created. Issues are identified by analyzing the concept maps created by participants and observing the participants while they are creating the map. The authors mention that they used this method to evaluate the usability of an API called Zoomable Object-Oriented Information Landscape (ZOIL) [25].

---

[1] https://www.mathworks.com

This method is a longitudinal usability evaluation methodology. It not only identifies issues that a programmer who is using the API for the first time face, but also capable of identifying usability issues that a user who has prior experience with the API would face [25]. Therefore, this methodology identifies issues related to aspects such as memorability and learnability that other methods do not identify. However, one of the limitations with this method is that it needs participants (i.e. programmers) to repeat the task for about 5 weeks, so it takes a long time for the whole evaluation [25].

*E. Automated evaluation*

Other than these methods, several tools have been introduced to evaluate the usability of APIs automatically [26]–[28]. de Souza and Bentolila [27] introduced a tool called Matrix, which evaluates the usability of an API by calculating its complexity. This tool assumes that usability is a function of complexity. It takes the API definition as the input and return complexity value for classes, packages and the overall API. The API concepts framework is another tool that evaluates the usability of APIs automatically [26]. This also works in a similar way to the method proposed by de Souza and Bentolila [27].

Comparing to other methodologies, automated usability evaluations are cheap, less time consuming and do not require experienced evaluators, test users or an implemented API [26], [27]. Furthermore, it can be easily integrated to the software development process. However, existing tools only consider complexity of the API when evaluating usability [26], [27]. Considering only complexity for evaluating the usability is not sufficient when it comes to security APIs. Because usability issues such as 'no clear documentation' and 'easy to misuse' are the major API usability issues that lead to introduce security vulnerabilities.

## III. COMPARISON OF API USABILITY EVALUATION METHODOLOGIES

Table I shows properties, strengths and weaknesses of usability evaluation methodologies that need to be considered when considering the effectiveness of a usability evaluation methodology at evaluating the usability of security APIs, and which of the above mentioned methodologies fulfil those properties.

Security vulnerabilities caused by usability issues of a security API occur when the programmer makes mistakes while using the API. Therefore, we argue that involving users of the API in the evaluation process is essential for evaluating the usability of security APIs. From Table I, we can see that 3 methodologies (User studies, API walk through method, API concept maps method) use programmers in the evaluation process. However, conducting user studies has been already tested in industry for evaluating usability of general APIs [11] and it identifies a wide range of usability issues exist in APIs [11], [15], [20]. Therefore, we propose user study based usability evaluation as the best approach to evaluate the usability of security APIs.

## IV. CONDUCTING USER STUDIES USING THE COGNITIVE DIMENSIONS FRAMEWORK

User study based usability evaluations are mainly conducted in four steps [11], [15], which are:

1) Designing tasks to employ programmers.
2) Recruiting participants and conducting the evaluation.
3) Collecting feedback from participants.
4) Identifying usability issues from participants' feedback

Following subsections describe how to carry out each of the steps in our proposed methodology.

*A. Step 1 : Designing Tasks To Employ Users*

User study based usability evaluations require participants to complete some tasks using the application under evaluation [11], [15], [17], [18]. In API usability evaluation, these tasks will be writing, debugging or reading a code that uses the API to achieve certain functionalities [15]. For example, in an API usability evaluation McLellan et al. [18] conducted, they used a task that has to read a programme that was developed using the API. When evaluating usability of a data persistence API, Piccioni et al. [15] used tasks that require participants to access a relational database using API features. The methodology we propose will use similar programming tasks for usability evaluation of security APIs.

*B. Step 2 : Recruiting Participants*

The participant programmers need to have experience in software development. It is important to select participants who are similar to programmers who will be using the API once it is developed. It is better to recruit programmers who are fluent in the corresponding programming language, so they will not come up with any issues due to their lack of proficiency in the programming language. Once participants are recruited, they will have to complete the tasks where they will have to use the API that is been evaluated. How many participants are required for this step is something that we need to identify by conducting experiments that employ this methodology.

*C. Step 3 : Collecting Feedback From The Participants*

Once a participant completes the task, evaluators should get the feedback about the API from the participant. Predefined questionnaires [11], [15], [22], semi structured interviews [18] and observation of participants while completing the task [17], [18] are some of the techniques that have been used to collect details about participants' experience. However, using a predefined questionnaire has its own advantages. The main advantage is that the participant does all the work, except questionnaire designing. Since the involvement of evaluators and API designers is minimum, the feedback collected from users will only reflect on their experience and opinion [29].

Microsoft uses this methodology where they use a predefined generic questionnaire based on the Cognitive Dimensions

TABLE I
PROPERTIES OF API USABILITY EVALUATION METHODOLOGIES THAT ARE RELEVANT TO SECURITY API.
✓: CORRESPONDING METHOD HAS THIS STRENGTH/WEAKNESS,
✗: CORRESPONDING METHOD DO NOT HAVE THIS STRENGTH/WEAKNESS,
- : THERE IS NO ENOUGH EVIDENCE TO COMMENT WHETHER OR NOT CORRESPONDING METHOD HAS THIS STRENGTH/WEAKNESS

| Property | Conducting user studies | Heuristic evaluation | API walkthrough | API concept maps method | Automated evaluation |
|---|---|---|---|---|---|
| Use programmers in the evaluation process | ✓ [11], [15] | ✗ [20], [21] | ✓ [24] | ✓ [25] | ✗ [26] |
| Ability to identify security related usability issues | ✓ [21] | ✗ [21] | - | - | ✗ [26] |
| Tested in the industry to general API usability evaluations | ✓ [11] | ✗ | ✓ [23], [24] | ✗ | ✗ |
| Low cost | ✗ [12], [20] | ✓ [20] | ✓ [23], [24] | - | ✓ [26] |
| Suitable to evaluate large APIs | ✗ | ✓ [20] | - | - | ✓ [26] |
| Less time consuming | ✗ [11] | ✓ [20] | ✓ [23], [24] | ✗ [25] | ✓ [26] |
| Identify issues with high severity | ✓ [21] | ✗ [21] | - | - | - |
| Longitudinal | ✗ [11], [15] | ✗ [20] | ✗ [23], [24] | ✓ [25] | ✗ [26] |
| Possibility to use early in the development process | ✗ [11], [15] | ✓ [20], [22] | ✓ [23], [24] | ✗ | ✓ [26] |
| Easy to embed into the software development process | ✗ | ✗ | ✓ [23], [24] | ✗ | ✓ [26] |
| Ability to identify large range of issues | ✓ [11], [15] | ✓ [20] | ✗ [24] | - | ✗ [26] |
| Evaluates API documentation | ✓ [11], [15] | ✓ [20], [22] | ✗ [24] | ✗ [25] | ✗ [26] |
| Do not require experienced evaluators | ✗ [11], [15] | ✗ [20] | ✗ [23], [24] | ✗ [25] | ✓ [26] |
| Do not require test users/evaluators | ✗ [11], [15] | ✓ [20] | ✗ [23], [24] | ✗ [25] | ✓ [26] |
| Do not require a finished API | ✗ [11], [15] | ✓ [20] | ✓ [23], [24] | ✓ [25] | ✓ [26] |

Framework to evaluate the usability of their APIs [11]. Using a generic questionnaire has several advantages over creating and using a specific questionnaire for each evaluation. It reduces the effort of evaluation since it eliminates the need to design a different questionnaire for each API [29]. Furthermore, since the API developer does not develop the questionnaire or select the dimensions to test, it will not be biased to the opinion of the developer [29]. Therefore, we recommend to use a predefined generic questionnaire based on the CDF to collect feedback about the usability of the evaluated API from participants who complete the task/tasks using the API.

However, the questionnaire used by Clarke [11] can not be used to evaluate the usability of security APIs, because there are more usability aspects that needs to be considered when evaluating security APIs that are not included in the Clarke's framework and questionnaire [3]–[5]. Wijayarathna et al. [3] proposed an enhanced version of the Clarke's framework and questionnaire by including security API related usability aspects proposed by Green and Smith [5], and Gorski and Iacono [4] to use in the usability evaluations of security APIs. It covers 15 cognitive dimensions that describe 15 usability aspects of security APIs, which are:

- Abstraction level
- Learning style
- Working framework
- Work-step unit
- Progressive evaluation
- Premature commitment
- Penetrability
- API elaboration
- API viscosity
- Consistency
- Role expressiveness
- Domain correspondence
- Hard to Misuse
- End User Protection
- Testability

Currently, this is the only such framework and questionnaire that is available to use in security API usability evaluations. Therefore, we are proposing to use this questionnaire for conducting user studies to evaluate the usability of security APIs.

*D. Step 4 : Identifying Usability Issues*

Since the CDF based questionnaire is an open-ended questionnaire, evaluators have to go through the answers provided by the participants and have to identify usability issues that participants encountered and reported.

Figure 1 shows a summary of the evaluation process we are proposing.

V. LIMITATIONS OF THE PROPOSED METHODOLOGY

We proposed conducting user studies using a cognitive dimensions based generic questionnaire as the best candidate for evaluating the usability of security APIs, because it has some qualities that we identified to be essential when evaluating the usability of security APIs. However, there are some limitations of this methodology also.

One of the major limitations of conducting user studies for usability evaluations of security APIs is the cost of conducting the evaluation [12], [20]. It can be costly with respect to time, resources and money. User studies require potential users of the API (i.e. programmers/ software developers) to use the API to complete some tasks. They should use the API for a while, so evaluators can evaluate the experience of the participant programmers. As potential users of an security API would be software developers/programmers, this can be somewhat costly [6], [7], [30].

Furthermore, using a predefined generic questionnaire to collect feedback can be time consuming compared to other feedback collection methods. Since the generic questionnaire that would be used is not designed for a particular API, participant programmers will have to answer questions that might not be relevant to the API that is being evaluated [29]. However, we propose this approach as it would eliminate the evaluator bias that the results would be affected from [29]. If the evaluator is involved in designing the API, they might miss some usability aspects that they would assume as not related to the particular API [29]. On the other hand, identifying usability issues from the feedback that participant give can be less time consuming compared to evaluations that would use feedback collection methods such as user observation and think aloud approach. Furthermore, the evaluators who conduct the experiment should have some expertise in usability and therefore can be a costly resource.

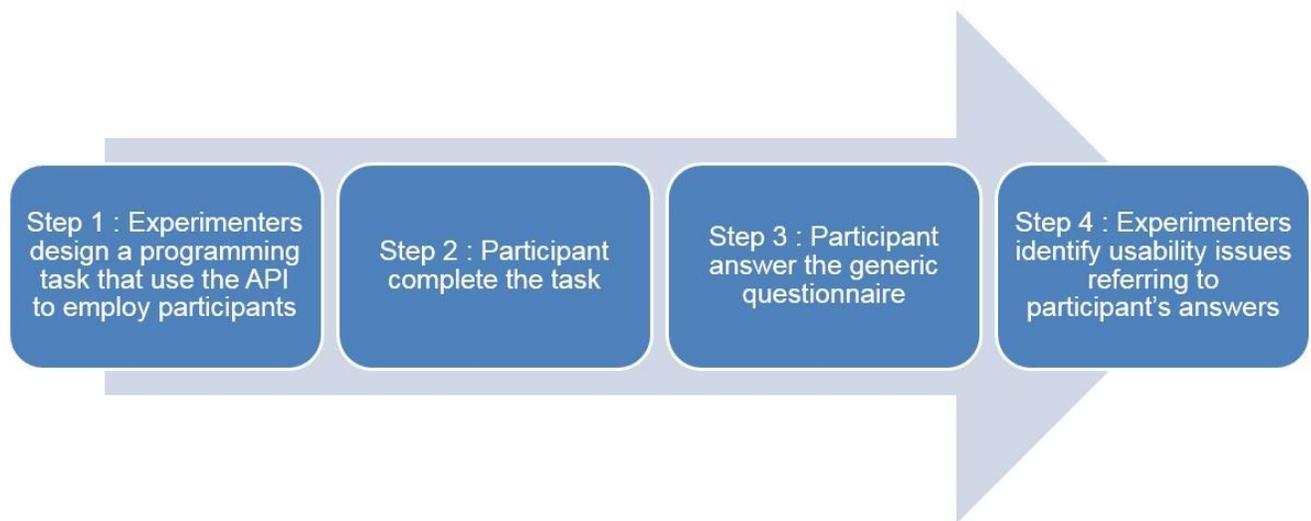

Fig. 1. Proposed process for evaluating the usability of security APIs

Due to the relatively high time requirement, the proposed methodology can be difficult to embed to the API development process. Furthermore, due to the high cost, conducting several iterations of the evaluation can be difficult.

Furthermore, this methodology might not be suitable to evaluate larger APIs as participant programmers have to use each component of the API in order to identify usability issues exist in them [6]. In a larger API, there can be a large number of components and participants will have to spend lot of time if they are to use all components of the API before giving their feedback. This would increase to cost of conducting the experiment.

In future work, we are attempting to improve the proposed methodology to overcome these identified weaknesses and reduce the cost of the evaluation process.

## VI. CONCLUSION

In this paper, we tried to identify a methodology to evaluate the usability of security APIs. We reviewed previous research done on evaluating the usability of general APIs and identified 5 methodologies that have been introduced and used to evaluate the usability of general APIs, which are,

- Conducting user studies.
- Heuristic evaluation.
- API walkthrough.
- API concept maps method.
- Automated evaluation.

By reviewing strengths and weaknesses of each of these methodologies, we proposed conducting user studies using a cognitive dimensions based generic questionnaire as the best candidate for evaluating the usability of security APIs. Finally, we discussed limitations of the proposed methodology that need to consider when improving the proposed methodology in future work.